\documentclass[journal=jacsat,manuscript=article]{achemso}

\usepackage[version=4]{mhchem}
\def\br{{\mathbf{r}}}
\newcommand{\jm}[1]{#1}
\usepackage{hyperref}
\usepackage{cleveref}
\usepackage{xr}
\graphicspath{{./figures/}}

\title{Density-Functionalized QM/MM Delivers Chemical Accuracy For Solvated Systems}

\author{Xin Chen}
\altaffiliation{Department of Physics, Rutgers University, Newark, NJ 07102}

\author{Jessica A. Martinez B.}
\altaffiliation{Department of Chemistry, Rutgers University, Newark, NJ 07102}

\author{Xuecheng Shao}
\altaffiliation{Department of Chemistry, Rutgers University, Newark, NJ 07102}
\altaffiliation{Key Laboratory of Material Simulation Methods and Software of Ministry of Education, College of Physics, Jilin University, Changchun 130012, China}

\author{Marc Riera}
\altaffiliation{Department of Chemistry, University of California-San Diego, San Diego, CA 92093}
\author{Francesco Paesani}
\altaffiliation{Department of Chemistry, University of California-San Diego, San Diego, CA 92093}
\author{Oliviero Andreussi}
\altaffiliation{Department of Chemistry, Boise State University, Boise, ID 83725}
\author{Michele Pavanello}
\email{m.pavanello@rutgers.edu}
\altaffiliation{Department of Chemistry, Rutgers University, Newark, NJ 07102}
\altaffiliation{Department of Physics, Rutgers University, Newark, NJ 07102}

\begin{document}

\maketitle

\begin{abstract}
We present a reformulation of QM/MM as a fully quantum mechanical theory of interacting subsystems, all treated at the level of density functional theory (DFT). For the MM subsystem, which lacks orbitals, we assign an ad hoc electron density and apply orbital-free DFT functionals to describe its quantum properties. The interaction between the QM and MM subsystems is also treated using orbital-free density functionals, accounting for Coulomb interactions, exchange, correlation, and Pauli repulsion. Consistency across QM and MM subsystems is ensured by employing data-driven, many-body MM force fields that faithfully represent DFT functionals. Applications to water-solvated systems demonstrate that this approach achieves unprecedented, very rapid convergence to chemical accuracy as the size of the QM subsystem increases. We validate the method with several pilot studies, including water bulk, water clusters (prism hexamer and pentamers), solvated glucose, a palladium aqua ion, and a wet monolayer of MoS$_2$.
\end{abstract}

\section*{Significance}
QM/MM is a powerful computational method used to model a critical, small portion of a complex molecular system—such as a protein’s active site--using quantum mechanics, while treating the surrounding environment with classical force fields. While QM/MM has advanced our understanding of enzymes and biological systems, it often struggles with accuracy, even when the QM and MM regions are not covalently bonded. A notable challenge is the slow convergence of system-environment interaction energies as the size of the QM region increases. This work demonstrates that incorporating quantum mechanics in the description of the MM subsystem leads to dramatically improved models. By assigning the MM subsystem a physically meaningful electron density, and using ab-initio density functionals for the QM-MM interaction (accounting for exchange, correlation, and Pauli repulsion), chemical accuracy in QM/MM models of aqueous solutions is achieved for the first time.

\section{Introduction}
QM/MM (standing for quantum mechanics/molecular mechanics) has revolutionized computational biochemistry \cite{Karplus2013NobelLecture}. Since the pioneering work of Honig and Karplus \cite{honig1971implications}, the combination of a quantum mechanical (QM) description for a subsystem with a classical point-charge description of its environment has led to major breakthroughs in fields such as enzymatics \cite{QMMMenzymes,KulikQMMM}, drug development \cite{drugQMMM,QMMM2023}, and materials design \cite{QMMMW}. Since its conception, methods handling the QM and MM subsystems have dramatically evolved. Today’s MM force fields can integrate data-driven potentials \cite{Paesani_polQMMM}, polarizable models \cite{GEM0,YKzhang_DrudeinQMMM,Thiel_DrudeinQMMM2,Benedetta2017_QMMMpol,Benedetta2020_QMMMpol,reinholdt2021fast,FlucCharDipol}, and even machine learning techniques \cite{zinovjev2023electrostatic,giese2024software}. QM methods have also evolved dramatically, from DFT methods to wavefunction theory methods routinely used in conjunction with QM/MM \cite{ratcliff2017challenges}.

The nature of the QM-MM interaction has also evolved. Initially, these were handled mechanically \cite{QMMMenzymes}, with Coulomb interactions calculated {\it a posteriori}, influencing only the forces and total energy, but without affecting the QM wavefunction or density. The advent of electrostatic embedding improved accuracy by incorporating MM partial charges directly into the QM Hamiltonian \cite{groenhof2013introduction}. Ultimately, mutual QM-MM polarization was achieved using polarizable force fields \cite{GEM0,Benedetta2020_QMMMpol}, a concept anticipated in early QM/MM work \cite{warshel1976theoretical}. 

The computational cost of QM/MM simulations is greatly reduced compared to fully quantum mechanical treatments. However, incorporating QM / polarizable-MM interactions in an efficient manner remains challenging. Methods based on judicious partitioning of the induction response have demonstrated excellent scalability \cite{reinholdt2021fast,lipparini2019general}. Additionally, algorithmic advances have been supported by steady progress in software development \cite{bondanza2024openmmpol,olsen2019mimic,cruzeiro2021open,pederson2024pydft,dziedzic2016tinktep,lu2023multiscale,dohn2017grid,giese2024software}. To extend QM/MM simulations to condensed phases, periodic boundary conditions (PBC) have been implemented. Ewald summation techniques are commonly used, and adaptations for molecular condensed phases \cite{QMMM-ewald-diss,QMMMuseChElPG,bonfrate2024analytic,pederson2022dft} and material systems \cite{laino2006efficient,laino2005efficient,dziedzic2016tinktep,dohn2017grid} are now widely available.

How accurate are QM/MM models? A common way to address this question is to evaluate the convergence of the results with respect to the size of the QM subsystem. Unfortunately, generally such a convergence is slow. Protein environments, for example, are exceptionally complex, and the search for effective ways to include relevant protein regions in QM/MM simulations continues \cite{brandt2022systematic,pedraza2020web,kulik2016large,karelina2017systematic}. Particularly challenging has been capturing charge transfer interactions on larger scales \cite{kulik2018large,kulik2016large}.

A slow convergence of the QM/MM setup has also affected those systems where partitioning in QM and MM subsystems involves no bond breaking. For example, water solvation \cite{perez2023effect}. Ironically, independent QM-only or MM-only treatments of liquid water can provide accurate results, but their combination in QM/MM workflows results in an overall reduced accuracy \cite{dziedzic2016tinktep,MHG_polQMMM}. Accurately modeling aqueous environments with QM/MM is essential due to the need to consider large water environments to properly account for the static and dynamic responses at water-material interfaces \cite{moscato2024unraveling,gomez2022multiple}. Therefore, representing these polarization effects in water bulk, which can extend for several nanometers, is crucial for capturing significant effects on the energetics of solvated species \cite{coons2018quantum,martinez2023physical}.

The culprit is the difficulty to accurately capture QM-MM interactions with a computationally efficient method. The MM subsystem is typically described using methods that largely (or completely) neglect its electronic structure. Point charges or, at times, point polarizable dipoles do not faithfully represent any electronic structure! The polarizable embedding method \cite{olsen2015polarizable} tackles this problem by dividing the MM subsystem into two regions: one near the QM subsystem is assigned a QM density derived from isolated fragment calculations, while the remaining MM atoms are treated using conventional point charge or dipole models. This embedding approach improves the accuracy of the QM Hamiltonian by capturing both electrostatic and non-electrostatic interactions, leading to better results than traditional QM/MM setups \cite{kvedaravi2023polarizable}. Similar methods, such as QM/ESP \cite{viquez2020exchange}, QM/GEM \cite{gokcan2018qm}, and QXD \cite{kuechler2015charge}, as well as approaches in density embedding \cite{tress2022employing} and many-body expansions \cite{gillan2013energy}, further explore these concepts. While attempts to account for the purely quantum mechanical Pauli repulsion within QM/MM have yielded mixed outcomes \cite{ben1998direct}, some methods address this by parametrizing the mechanical embedding interaction energy without introducing new terms in the QM Hamiltonian \cite{pauli_exchange1, chemRev_Pauli,hrvsak2017optimization}.

Thus, our approach in this work is to treat QM and MM subsystems on a more equal basis, aiming to reduce the impact of an imbalanced QM-MM interface and an imbalanced treatment of the internal energy of QM and MM subsystems. We propose ``density-functionalizing" the MM subsystem, assigning it an electron density such that it can be handled like an electronic subsystem within the rigorous framework of subsystem DFT \cite{krishtal2015subsystem,jacob2014subsystem,wesolowski2015frozen,mi2023orbital,jacob2024subsystem}. This standardization of QM and MM subsystems allows for the use of first-principles density functionals for evaluating the QM-MM interaction, inherently capturing all relevant physical effects, such as exchange, correlation, Pauli repulsion, electrostatics, and charge penetration. The next section details the theoretical framework for this density-functionalized QM/MM approach, with additional, less critical details provided in the supplementary materials. \cite{epaps}

\section{A density-functionalization of QM/MM}

The central idea is to assign an electron density to both the QM subsystem, $\rho_{QM}(\br)$, and to the MM subsystem, $\rho_{MM}(\br)$, with the total electron density given by their sum and the energy functional borrowed from rigorous subsystem DFT \cite{krishtal2015subsystem,jacob2014subsystem,wesolowski2015frozen} (sDFT, hereafter),
\begin{align}
    \label{eq:den}
    \rho(\br) &= \rho_{QM}(\br) + \rho_{MM}(\br),\\
    \label{eq:ene}
    E[\rho_{QM},\rho_{MM}] &= E[\rho_{QM}] + E[\rho_{MM}] + E^{nad}[\rho_{QM},\rho_{MM}].
\end{align}

Although formally the electronic energy is strictly a density functional, in practice, the external potential (electron-nuclear attraction) is known ahead of time and is thus specified for the QM subsystem, $v_{QM}(\br)$, and for the MM subsystem, $v_{MM}(\br)$, such that the additive part of the energy is given by (disregarding for the time being the nuclear-nuclear repulsion)
\begin{align}
    \label{eq:addQM}
    E[\rho_{QM}] &= T_s[\rho_{QM}] + E_H[\rho_{QM}] + E_{xc}[\rho_{QM}] + \int v_{QM}(\br)\rho_{QM}(\br) d\br, \\
    \label{eq:addMM}
    E[\rho_{MM}] &= T_s[\rho_{MM}] + E_H[\rho_{MM}] + E_{xc}[\rho_{MM}] + \int v_{MM}(\br)\rho_{MM}(\br) d\br,
\end{align}
where $T_s$, $E_H$ and $E_{xc}$ are the noninteracting kinetic energy, the classical electron-electron repulsion (Hartree) and the exchange-correlation functionals, respectively.
The nonadditive energy is thus given by
\begin{align}
    \label{eq:nad}
    E^{nad}[\rho_{QM},\rho_{MM}] &= T_s^{nad}[\rho_{QM},\rho_{MM}]+E_{xc}^{nad}[\rho_{QM},\rho_{MM}]+ \\
    \nonumber
    &+ \int \frac{\rho_{QM}(\br)\rho_{MM}(\br)}{|\br-\br^\prime|}d\br d\br^\prime + \int \rho_{MM}(\br)v_{QM}(\br)d\br + \int \rho_{QM}(\br)v_{MM}(\br)d\br
\end{align}
where $T_s^{nad}[\rho_{QM},\rho_{MM}]=T_s[\rho_{QM}+\rho_{MM}] - T_s[\rho_{QM}]-T_s[\rho_{MM}]$ and equivalently for $E_{xc}^{nad}$.

The equations above can be exploited for a variational minimization of the energy functional with respect to variations in both $\rho_{QM}$ and $\rho_{MM}$, provided that suitable approximations for the relevant density functionals are available. 
When both subsystems are treated at the Kohn-Sham DFT level, this approach yields sDFT, which was found to be accurate in the limit of weak inter-subsystem interactions. This is the case for systems such as water molecules in liquid water \cite{genova2016avoiding} or CO$_2$ molecules in fluid CO$_2$ \cite{mi2019ab}. In fact, sub-1 kcal/mol accuracy is now routinely achieved in sDFT calculations either with Kohn-Sham subsystems \cite{shao2021gga,mi2019nonlocal,schluns2015subsystem}, orbital-free subsystems \cite{shao2022density}, and proper multiscale simulations \cite{shao2022adaptive,chen2024unraveling,schmitt2021density,focke2023coupled}. Inter-subsystem interactions involving hydrogen bonds, are the focus of this work and are particularly well described by the nonadditive PBE exchange-correlation and revAPBEk nonadditive kinetic energy functionals \cite{shao2021gga} (and in general many GGA nonadditive functionals \cite{kevorkyants2006interaction}).

In a QM/MM framework, Eq.\ (\ref{eq:addMM}) is usually replaced by a classical force-field expression, which involves electrostatic interactions between atom-centered point charges and possibly atom-centered polarizable dipoles, as well as (typically empirical) expressions for short range dispersion-repulsion interactions and bonded terms. In common implementations of QM/MM approaches, the tool associated with the classical component is responsible for the calculation of this energy term. However, the electrostatic multipoles present in the MM force-field are also involved in the electrostatic terms of the QM-MM interaction energy term in Eq.\ (\ref{eq:nad}). Usually, ad-hoc corrections to describe short-range dispersion-repulsion interactions and avoid unphysical behaviors when the QM and MM subsystems are too close are introduced to approximate the first two terms of Eq.\ (\ref{eq:nad}). 
Once the analytic dependence of Eq.\ (\ref{eq:ene}) on the atomic positions of the MM subsystem is defined, the functional derivative of Eq.\ (\ref{eq:ene}) with respect to $\rho_{QM}(\br)$ allows the optimization of the QM subsystem via a self-consistent field (SCF) approach.

Point-charges and point-dipoles introduced in the definition of the MM part of the system can be tuned to fit empirical results for the MM system, or they can be optimized to reproduce properties connected to the electronic density of the MM components computed from first principles (e.g., binding energies or the behavior of the electrostatic potential). Because the permanent charges in MM force fields are independent of the geometry of the system and of its surrounding environment, atomic polarizabilities may be included in the force field in order to capture the system’s electrostatic response to external fields. The resulting polarizable force fields will, therefore, respond to the non-additive term in Eq.\ (\ref{eq:nad}) as it contributes to the induced dipoles in the MM subsystem. This results in mutually polarized QM and MM subsystems. 

The typical approximations in the above approach lead to significant inaccuracies when the boundary between QM and MM regions varies systematically. Drawing from existing QM/MM methods that account for electron densities in the MM subsystem \cite{olsen2015polarizable, GEM-AMOEBA} and the successes of the sDFT framework, we hypothesize that a QM/MM framework based on Eqs.\ (\ref{eq:ene}–\ref{eq:nad}) can accurately model solute-solvent systems. This extends to weakly interacting QM and MM subsystems, provided consistent forward and backward mappings between first-principles electronic densities and force field multipoles are established.


For the forward mapping between electronic-structure calculations and accurate classical electrostatics, the many-body polarizable approach by Paesani and collaborators showed that it is possible to effectively describe statistical properties of bulk systems by carefully parametrizing a classical force-field on a large database of accurate few-body first-principles simulations \cite{MB_pol1,MB_pol2}. While the initial developments of MB-Pol focused on high-accuracy coupled-cluster simulations of water dimers and trimers,  the approach can be applied to DFT-based calculations \cite{MBDFT} giving rise to, e.g., the MB-PBE force field. The resulting force-fields describe electrostatic interactions in terms of atomic charges and polarizabilities. As in similar approaches in the literature, short-range corrections are introduced to avoid self-polarization and the unphysical divergence of the induced dipoles when different polarizable systems are too close to each other. In this work, we hypothesize that using an MM subsystem whose electrostatic interactions are fully consistent with the level of theory of the QM component will allow the seamless convergence of QM-MM calculations as the boundary between the two regions is varied.

For the backward mapping between the classical force-field used in the MM region and an effective electronic density to use in the density functionalized QM-MM interaction term, we introduce the following approach. We treat each atomic nuclei and core electrons using pseudopotentials. To keep the approach computationally efficient for large MM subsystems, we use the local part of the ultrasoft pseudopotentials from Garrity and Vanderbilt (GBRV) \cite{GBRV}, which are designed for fast, high-throughput simulations that require low plane wave cutoffs. This assignment can be efficiently handled in a computationally linear scaling manner using the particle-mesh Ewald method \cite{shao2016n}. For the valence electrons, we convert the point-like charge multipoles computed and used by the classical force-field into a smooth electronic charge density. 
Namely, for each atom $i$, the valence electron density is represented by a Gaussian centered at the ion’s position, $\mathbf{R}_i$, with an adjustable width $\sigma_i$: $\rho_{q_i}(\mathbf{r}) = (N_i - q_i) g_{\sigma_i}(\mathbf{r} - \mathbf{R}_i)$, where $g_{\sigma_i}$ is a normalized Gaussian and the prefactor is crucial for accurately representing the ion’s permanent charges, $q_i$, while accounting for the isolated atom’s number of valence electrons, $N_i$. If the force-field involves higher multipoles in the description of its electrostatic interactions, they can be included in the reconstructed electron density by using derivatives of the normalized Gaussian. In particular, we can map a point-like dipole into a corresponding smooth density using the gradient of a Gaussian function as $\rho_{\mu_j}(\mathbf{r}) = -\vec{\mu_j} \cdot \vec{\nabla} g_{\sigma_j}(\mathbf{r} - \mathbf{R}_j)$, where $\vec{\mu_j}$ is the induced dipole at the dipole site $j$. Similarly, higher-order multipoles can be mapped using higher-order derivatives. The total valence electron density becomes 
\begin{equation}
    \label{eq:MMden}
    \rho_{MM}(\br) = \sum_{i\in \text{MM charges}} \rho_{q_i}(\br) + \sum_{j\in \text{MM dipoles}} \rho_{\mu_j}(\br).
\end{equation}
We stress that the proposed formulation relies on a set of element-specific widths (the $\sigma_i$) that can significantly affect the final shape of the reconstructed electronic density (both permanent charges and polarization density due to the induced dipoles). 

For the forward mapping, we hypothesize that using the Gaussian widths that provide the best match between QM/MM vs QM/QM (sDFT) interaction energies will allow for the accurate description of the mutual polarization effects and the seamless convergence of QM/MM calculations as a function of an increasing size of the QM subsystem. 

The functional derivatives of the non-additive interaction energy with respect to the QM or MM subsystem densities yields the embedding potentials
\begin{align}
    \label{eq:emb_MM}
    v_{emb}^{QM}(\br) &= \frac{\delta E^{nad}[\rho_{QM},\rho_{MM}]}{\delta \rho_{QM}(\br)} = v_{MM}(\br) + \int d\br^\prime \frac{\rho_{MM}(\br^\prime)}{|\br-\br^\prime|} + \frac{\delta T_s^{nad}}{\delta \rho_{QM}(\br)} + \frac{\delta E_{xc}^{nad}}{\delta \rho_{QM}(\br)},\\
    \label{eq:emb_QM}
    v_{emb}^{MM}(\br) &= \frac{\delta E^{nad}[\rho_{QM},\rho_{MM}]}{\delta \rho_{MM}(\br)} = v_{QM}(\br) + \int d\br^\prime \frac{\rho_{QM}(\br^\prime)}{|\br-\br^\prime|} + \frac{\delta T_s^{nad}}{\delta \rho_{MM}(\br)} + \frac{\delta E_{xc}^{nad}}{\delta \rho_{MM}(\br)},
\end{align}
that can be used to optimize the QM or MM degrees of freedom. In particular, Eq.\ (\ref{eq:emb_QM}) enters the Hamiltonian of the QM subsystem(s) at each SCF step and it is used for the calculation of the ground state QM electronic densities as well as for the optimization of QM atomic positions. The sDFT framework and the implementation of this approach in the eDFTpy software \cite{shao2021edftpy} allows for the coupling of the QM/MM embedding potential into multiple QM subsystems.

Eq.\ (\ref{eq:emb_MM}) instead can be used by the MM engine to compute the QM effects on the induced dipoles and on the interatomic forces. For the former, the gradient of the MM embedding potential needs to be added to the classical electric field used to compute induced dipoles. While all the terms in Eq.\ (\ref{eq:emb_MM}) would ideally be included in the embedding field that polarizes the MM component, the current definition of atomic polarizabilities within polarizable force fields is only meaningful for a classical description of long-range polarization effects. Thus, in the current implementation, we only keep the first two terms, which are related to classical electrostatic interactions, and we neglect the effect of the non-additive kinetic and exchange-correlation terms, which are more relevant for short-range interactions. 
\begin{figure}[htp]
  \centering
  \includegraphics[width=0.8\linewidth]{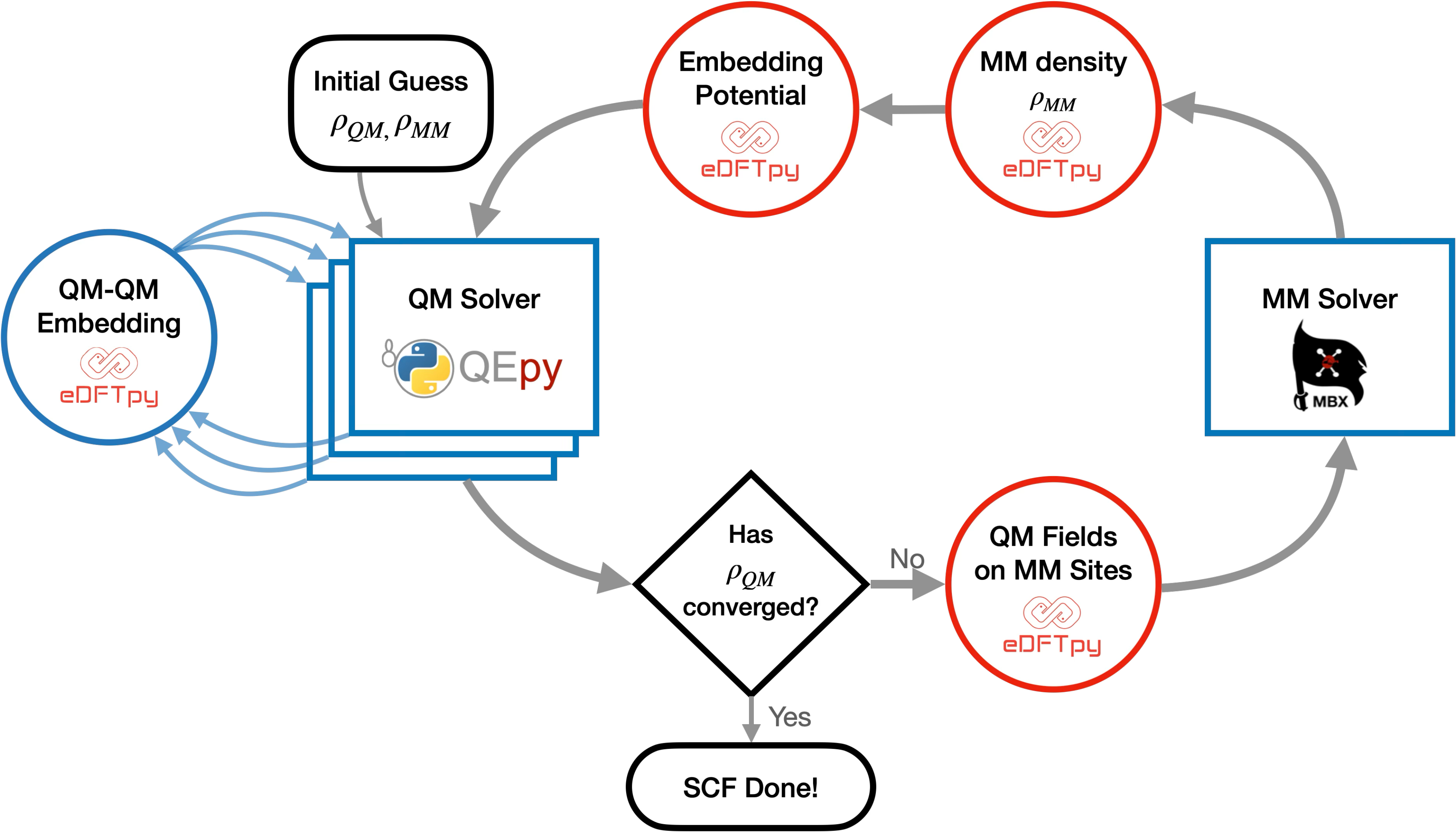}
  \caption{Workflow of the QM/MM method with emphasis on the software implementation. See details in section \ref{sec:implementation}.}
  \label{fig:workflow}
\end{figure}

The evaluation of the QM additive energy in \eqref{eq:addQM} can utilize the sDFT implementation of eDFTpy with either a single QM subsystem or multiple QM subsystems. When $N_S$ QM subsystems are involved, the QM electron density is represented as the sum of contributions from each subsystem, and the QM energy is decomposed into additive and nonadditive terms, following the structure of \eqref{eq:ene}. Namely,
\begin{align}
    \label{eq:qmsub}
    \rho_{QM}(\br) &= \sum_{I=1}^{N_S}\rho_I(\br),\\
      E[\rho_{QM}] &= \sum_{I=1}^{N_S} E[\rho_I] + E^{nad}[\{ \rho_I \}].
\end{align}
In the results Section \ref{sec:results}, we will refer to a one-subsystem treatment of the QM region as fragmentation type 1 (or Frag.\ 1) while a many-subsystem treatment as fragmentation type 2 (or Frag.\ 2).

\section{\label{sec:implementation}Details of the implementation}
We now discuss the details of the implementation of the QM/MM algorithm in the eDFTpy software \cite{shao2021edftpy}. Figure \ref{fig:workflow} illustrates the QM/MM workflow of the SCF cycle in the presence of a MM subsystem. It begins with initial guesses for the MM and QM density and wavefunctions.  In eDFTpy, we use QEpy \cite{shao2021qepy} as QM solver (a python implementation of Quantum ESPRESSO 7.2 \cite{QE-2017}) and a python interface to MBX as the MM solver \cite{MBX2023}.  The initial guess density for the QM subsystem is usually taken from the sum of atomic densities from the pseudopotential files. For the MM subsystem, a valence electron density consistent with the value of the permanent charges (which are fixed, independent of geometry) is used. 

eDFTpy then directs QEpy, the QM solver, to solve for the electronic structure of the QM system. This is done either by a single QEpy instance when a single QM subsystem is considered, or by multiple QEpy instances when several QM subsystems are considered. The QM subsystems require the computation of QM-in-QM embedding potentials which are handled also by eDFTpy as indicated by the blue arrows connecting the QEpy solvers to the eDFTpy circle in the figure. The QM electrostatic field formally given by the negative gradient of the MM embedding potential in Eq.\ (\ref{eq:emb_MM}), $-\vec\nabla v_{emb}^{MM}(\br)$ is then represented on the MM polarizable dipole sites. For large systems, representing the QM electric field on the MM sites is a task of non-negligible computational complexity \cite{s2016multipole,ferre2002approximate,laino2005efficient,laino2006efficient}.  The QM electric field is then represented on the grid point closest to each of the MM polarizable dipole sites. This is accurate enough given the exceptional smoothness of the QM field in the MM region \cite{laino2005efficient} and saves the inconvenient step of splining the field on the exact position of the ions. Spline would require an \texttt{MPI\_GATHER} operation which carries a high wall-time cost. In eDFTpy, the MM cell is spanned by a grid fine enough to properly represent the ionic pseudopotentials centered on each QM ion and MM site. In our simulations, we use a 100 Ha cutoff for this grid, which is fine enough to produce real-space grid points spaced by about 0.23 $a_0$.

Once the electric field is represented on the MM sites, the MM solver is tasked with solving the Coulomb problem in the MM cell to yield the MM polarizable dipoles. The MM density is then built using Eq.\ (\ref{eq:MMden}). Having the MM density and the density of all QM subsystems allows eDFTpy to compute the required embedding potentials for each subsystem, including the MM subsystem. The SCF cycles then continue until convergence is achieved, which is calibrated on the convergence of the QM density. At convergence, QM density and MM dipoles are fully mutually polarized.

In the eDFTpy implementation we take full advantage of plane wave reduction techniques that were developed for sDFT\cite{genova2017eqe,mi2021eqe,shao2022adaptive,shao2022density,genova2015exploiting}. Specifically, MM and QM subsystems are represented by different simulation cells, grids and thus plane wave basis sets. The QM simulation cell is smaller compared to the physical cell which coincides with the MM cell. Coulomb fields and other long-ranged energy functionals are evaluated on the large, physical cell. Employing such a multi-cell/grid approach is crucial in the context of QM/MM simulations as the MM subsystem is usually dramatically more extended than any of the QM subsystems. In the supplementary materials \cite{epaps} we further discuss our parallelization strategy.

\begin{figure}
  \centering
  \includegraphics[width=0.65\linewidth]{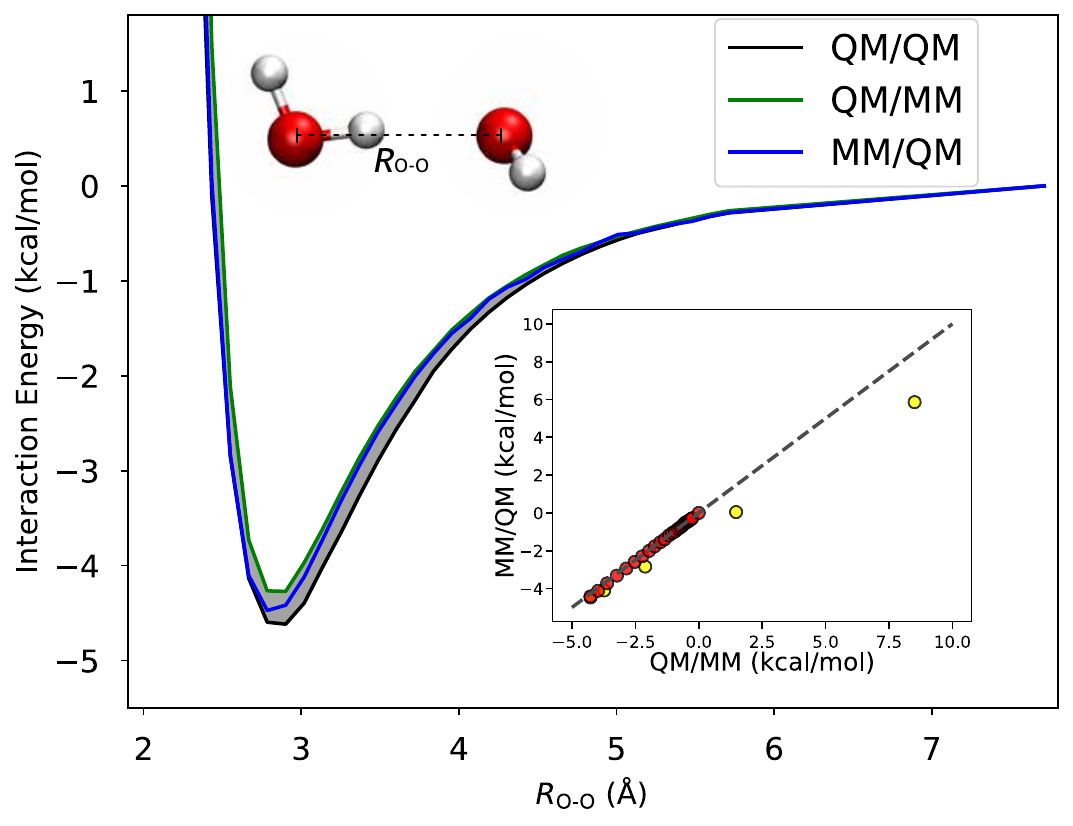}
  \caption{\label{fig:dimer} Water dimer energy curve (structure shown) for O--O distances ranging from 2.3 to 7.7 \AA. The black line represents the QM/QM result, where both the donor and acceptor monomers are treated at the QM level. The shaded area represents the deviation of the QM/MM result from QM/QM. Green and blue lines correspond to the QM hydrogen bond donor (QM/MM) and acceptor (MM/QM), respectively. The inset shows a correlation plot of QM/MM and MM/QM interaction energies. Yellow markers represent geometries in the repulsive region of the curve ($R_\text{O-O}<2.9$ \AA), while red markers correspond to those in the attractive region ($R_\text{O-O}>2.9$ \AA).}
\end{figure}

The current implementation also features analytic energy gradients for the atoms in the QM region. In the supplementary materials we devote a section to the implementation of the forces and results are shown in Tables \ref{sup-tab-forces} and \ref{sup-tab-gluc_opt}.

\section{Computational Details}
\subsection{Pseudopotentials, Density Functionals and Plane Wave Cutoffs}
GBRV pseudopotentials \cite{GBRV} are used for all elements considered of both the QM and MM subsystems. The PBE exchange-correlation functional \cite{PBEfunc} and the revAPBEk noninteracting kinetic energy functional \cite{laricchia2011generalized} are employed for approximating additive $E_{xc}$ and nonadditive $E_{xc}$ and $T_s$ functionals. All calculations include Grimme's D3 correction \cite{grimme2012supramolecular}. We choose plane wave basis sets for the QM subsystems with a cutoff energy of 20 Ha for wavefunctions and 200 Ha for charge density and potential unless otherwise stated (see supplementary Table \ref{sup-tab-st1}). The energy convergence threshold for the SCF was set to $10^{-8}$ Ha/atom. The Brillouin zone was sampled by a $5\times 5 \times 1$ k-point mesh for MoS$_2$ and the $\Gamma$ point for all other calculations. MM calculations were conducted using the MBX package\cite{MBX2023} using the MB-PBE and MB-Pol water models, which were developed to quantitatively reproduce PBE or CCSD(T) water \cite{MBDFT,babin2013development}, respectively.

All inputs/output files, Jupyter notebooks needed to analyze the data and reproduce the figures in this work, as well as links to tagged versions of the software (MBX, eDFTpy, QEpy, and DFTpy) used for the simulations, are available as reported in the supplementary materials \cite{epaps}.

\subsection{Parameters Defining the MM Density}

As described in the sections above, the proposed framework relies on the conversion of classical permanent charges and polarizable dipole sites into a smooth electronic density. For each atomic charge and polarizable dipole, this involves the fit of the width, $\sigma$, of the corresponding normalized Gaussian functions that contribute to the expansion in Eq.\ (\ref{eq:MMden}). In our applications to solvated systems, MM sites only involve oxygen and hydrogen atoms, for a total of 4 parameters that need to be fitted. 

We also considered an additional MM-induced dipole self-energy correction. Although the induced point-dipole self-energy is given by $\frac{1}{2}\mu^2 \alpha^{-1}$ \cite{Bottcher1973iii}, where $\alpha$ is the isotropic dipole polarizability, we recognize that the dipoles considered in our work are not point dipoles, e.g., their charge density overlaps with the QM density at the QM/MM interface. Thus, we added an additional term to the self-energy for each site equal to $k^{SE}_{i}|\vec\mu - \vec\mu^\prime|^2$, where $\vec\mu^\prime$ is the induced dipole at the same site when only the MM subsystem is considered, and $k^{SE}_i$ are element-dependent proportionality constants. Additional details about this correction and about how the permanent charge density was generated for the MB-PBE and MB-Pol force fields (which use the so-called M-site) are available in the supplementary materials document \cite{epaps}.

The parameters defining the MM density were fitted so as to reproduce the QM-QM interaction energies for a single water molecule in bulk water. Ten snapshots of 64-water molecule cubic systems were taken from Ref.\ \citenum{kumar2017cooperation}. These provided 640 water-bulk interaction energies. More details about this system will be given in the results section. The final values of the parameters for both MB-PBE and MB-Pol force fields are listed in Table \ref{sup-tab-st1}. 

The use of bulk simulations as a reference for the parametrization of the density functionalization approach provides a robust and general strategy that reduces the need for application-specific benchmarks. Results on small water clusters and solvation effects on molecules and materials (reported in the following) highlight the transferability of the obtained parameters. 

\section{\label{sec:results}Results and Discussion}

\subsection{Water Dimer}

\begin{figure}[t]
  \centering
  \includegraphics[width=0.7\linewidth]{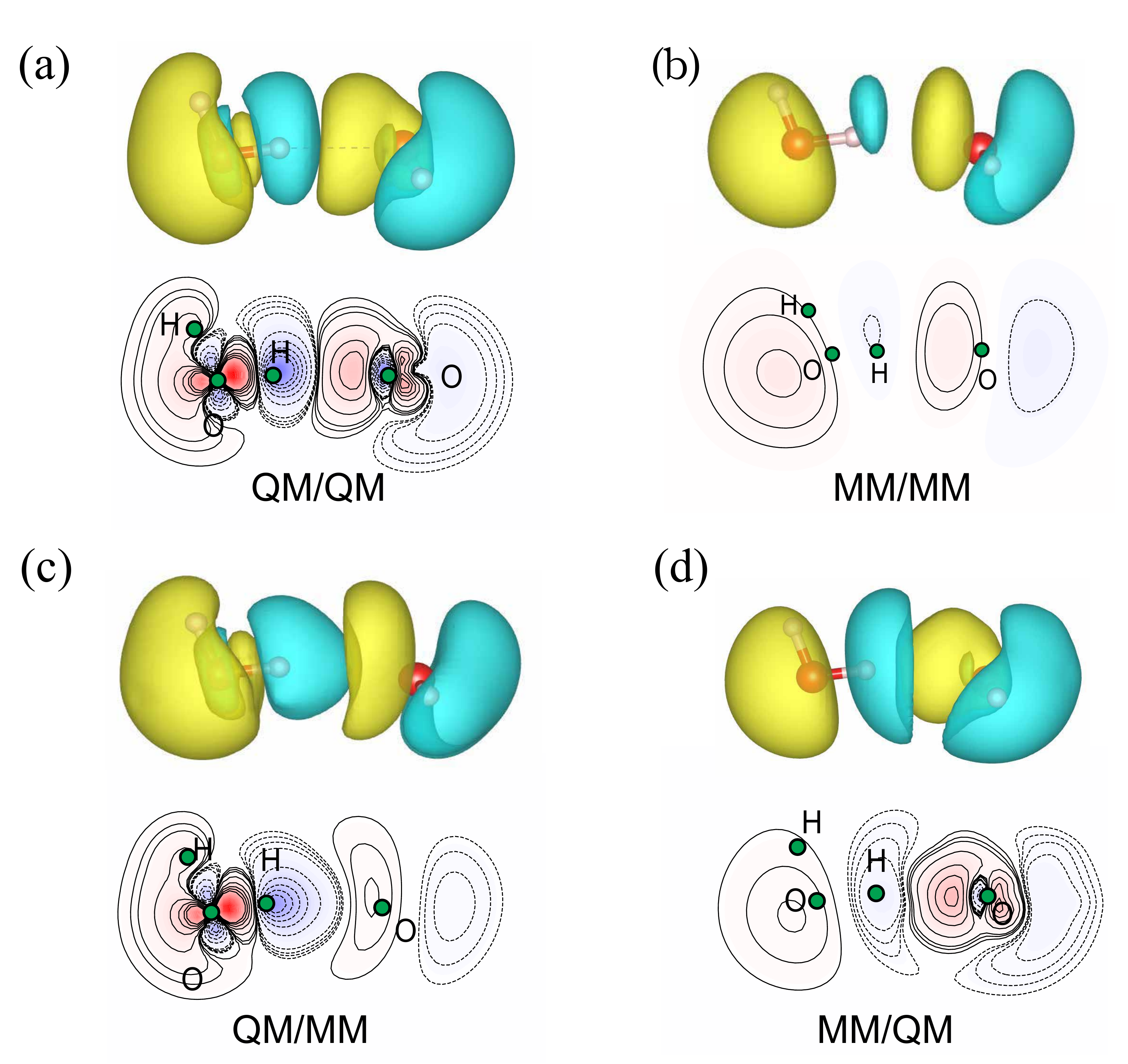}
  \caption{Polarization density, defined as $\rho(\br) - \rho_\text{iso}(\br)$, where $\rho_\text{iso}$ is the sum of the electron densities of the isolated water monomers, for the water dimer at the equilibrium O--O distance. In each panel, we present isosurfaces (top) and contour plots (bottom) generated with a cutoff of $\pm 0.0007\, e\cdot a_0^{-3}$.}
  \label{fig:dimeredd}
\end{figure}

We start by comparing the QM/MM potential energy curves shown in Figure \ref{fig:dimer} and the polarization density depicted in Figure \ref{fig:dimeredd} for water dimers. The dimer structures, sourced from Ref.\ \citenum{waterdimerCCSD}, were placed in a 20-\AA\ cubic simulation cell and evaluated against benchmark QM/QM simulations performed using sDFT.


The key aspect of this system lies in its asymmetry, as one monomer is a hydrogen bond donor, and the other is an acceptor. Even though both QM and MM monomers are modeled by PBE (the MM subsystem is described by the MB-PBE model), the nature of the QM-MM interface is dramatically different whether one considers a QM hydrogen bond donor (QM/MM) or acceptor (MM/QM). Despite this, the curves are remarkably similar. The QM/MM and MM/QM minima (at 2.90 \AA \ O--O distance) are less than 0.15 kcal/mol away from each other, and both deviate from the reference by a similar measure. These results show that the QM/MM interface is extremely well characterized by the nonadditive functionals. Crucially, the repulsive part of the curve is also well reproduced. 
Our $\sim$0.2 kcal/mol error compares well with the $\sim$0.6 kcal/mol error for the dimer reported in Ref.\ \citenum{MHG_polQMMM} for their QM/MM (AMOEBA) method.

In the inset of Figure \ref{fig:dimer}, we report the correlation between the interaction energies of the QM/MM and MM/QM systems, showing that they are in extremely good agreement. The interaction energies in the attractive part of the curve follow the ideal trend, i.e., the points sit on the diagonal line. However, those in the repulsive part (yellow color coded markers) show a slightly deviated trend. A behavior indicative of an asymmetry in accounting for charge penetration effects of the H$_\text{QM}$-O$_\text{MM}$ vs H$_\text{MM}$-O$_\text{QM}$.
To ensure our results are not affected by possible artifacts due to the use of PBCs, we recalculated the interaction energy minima in a larger simulation cell with lattice constant $a=50$ \AA\ finding a shift of less than 10$^{-3}$ kcal/mol.

Figure \ref{fig:dimeredd} shows the polarization density (defined in the caption) of the water dimer system. It is clear that the MM polarization is only qualitatively similar to the QM polarization. Specifically, comparing the QM/QM (panel a) with the MM/MM (panel b) shows that the latter misses several potentially important details near the ion's core and along the O--O line. These are well-known deficiencies in the electrostatic response of dipole-only polarizable force fields \cite{holt2010nemo,holt2008inclusion}. Interestingly, the QM/MM treatment improves the polarization of both monomers. Panels (c) and (d) show that the MM molecule polarization gains features that are prominent in the QM response of that monomer but that are absent (or weakly present) in the MM/MM case. This shows that the handling of the QM/MM interface by our method is accurate not only by the interaction energy measure (recall Figures \ref{fig:dimer}) but also by the much more stringent test of subsystem mutual polarization.

\subsection{Water Hexamer}


Hexamer water structures play a crucial role in quantum chemistry due to their complex hydrogen-bond topologies. Accurately predicting their relative energies is often seen as a benchmark for a model's ability to represent water across its various stable phases \cite{{prismhex}}. Here, following Ref.\ \citenum{Paesani_polQMMM}, we focus on the most stable hexamer, the prism hexamer, to examine the effect of the QM-MM boundary. In the prism hexamer all water monomers act as both hydrogen bond donors and acceptors, with either 1/2 or 2/1 accepted/donated hydrogen bonds. 

The root mean square errors (RMSEs) of the computed interaction energies defined as the energy of the hexamer minus the energy of the 6 isolated water molecules are collected in Figures \ref{sup-fig:hexamer} and \ref{sup-fig:mbpol-hex} for the MM treated with MB-PBE and MB-Pol, respectively. When multiple water molecules are treated at the QM level, the sDFT framework allows flexibility in how the QM waters are grouped: they can be combined into a single subsystem (Frag.\ 1 in the figure) or divided into separate subsystems (Frag.\ 2). For each partition type with $k$ QM or MM water molecules, there are $\binom{6}{k}$ possible members of the partition (i.e., ways to split the hexamer into QM and MM subsystems).

Ideally, all members of each partition should yield the same interaction energy. Therefore, a larger RMSE for the predicted interaction energy indicates a lower accuracy of the method. Recognizing that no practical fragmentation method is perfect, we find a relation between the RMSE for the $k$-th partition ($\sigma_{\rm QM/MM}(k)$) and the square root of the number of members in the partition \cite{ray905problem} ($\sqrt{\binom{6}{k}}$). This relationship is confirmed by $R^2$ values of 0.99 for all QM/MM and fragmentation methods considered (see Figure \ref{sup-fig:sf1}). The slope of this linear relationship is proportional to the average error per QM/MM boundary in the partitions (error = slope $\cdot \sqrt{\frac{\pi}{2}}$), yielding a predicted error per boundary ranging between 0.2 and 0.3 kcal/mol for the methods considered. This error is consistent with the results for water dimers and, as we will see in the next section, also aligns with the RMSEs for the pentamer clusters extracted from bulk liquid structures.

\subsection{Bulk and First Solvation Shell Water Environment}\label{bulk_sect}

To further assess our method, we consider the dipole moment and the molecule-environment interaction energy of water molecules embedded in an MM environment of bulk liquid water. The benchmark is once again sDFT. Dipole moments of embedded molecules are  accessible in sDFT, calculated with the subsystem electron density. This procedure was found to be accurate for a variety of embedded molecular species \cite{krishtal2015subsystem,krishtal2016revealing,kumar2017cooperation,luber2014local}.  We consider 10 snapshots of an ab-initio dynamics trajectory of 64 independent water molecules in a cubic cell of lattice constant 12.42 \AA\ that was presented elsewhere\cite{genova2016avoiding,kumar2017cooperation} (also available in the supplementary materials \cite{epaps}). We use the nomenclature $n/m$ to denote a simulation where $n$ molecules are treated at the QM level and $m$ are treated at the MM level. In all cases, the reported quantity is the interaction of a single water molecule with the environment. For example, 1/63 means there are 63 MM water molecules and one QM water, for a total of 64 water molecules. We use sDFT to generate benchmark values for  the water-environment interaction energy. Within a sDFT framework, one may treat all 63 environment molecules as a single subsystem or as 63 coupled subsystems (for the 1/4 system we only consider 4 environment water molecules). Owing to the accuracy of sDFT for bulk water simulations \cite{genova2016avoiding,martinez2023physical}, we employed both approaches, which resulted in very similar trends, as reported in Figure \ref{sup-fig:int_ener_63_subs} of the supplementary materials. For the discussion in the following, we consider the benchmark where all 63 molecules in the environment are grouped in a single subsystem. Leveraging the available bulk configurations, inspired by Ref.\ \citenum{MHG_polQMMM}, we also considered the performance of our QM-MM approach on ``first shell'' structures, where each water molecule of the bulk is embedded only by the nearest 4 water molecules of the environment, giving rise to a collection of water pentamer structures (denoted by 1/4). We also test our QM/MM scheme on a partition ``minimal solvation'' where each water molecule and the nearest 4 water molecules of the environment are included in the QM subsystem and the remaining 59 water molecules are kept in the MM subsystem (system 5/59). Because in each snapshot there are 64 water molecules and there are 10 snapshots, 640 total structures / data points are available. 

\begin{figure}
  \centering
  \includegraphics[width=0.78\linewidth]{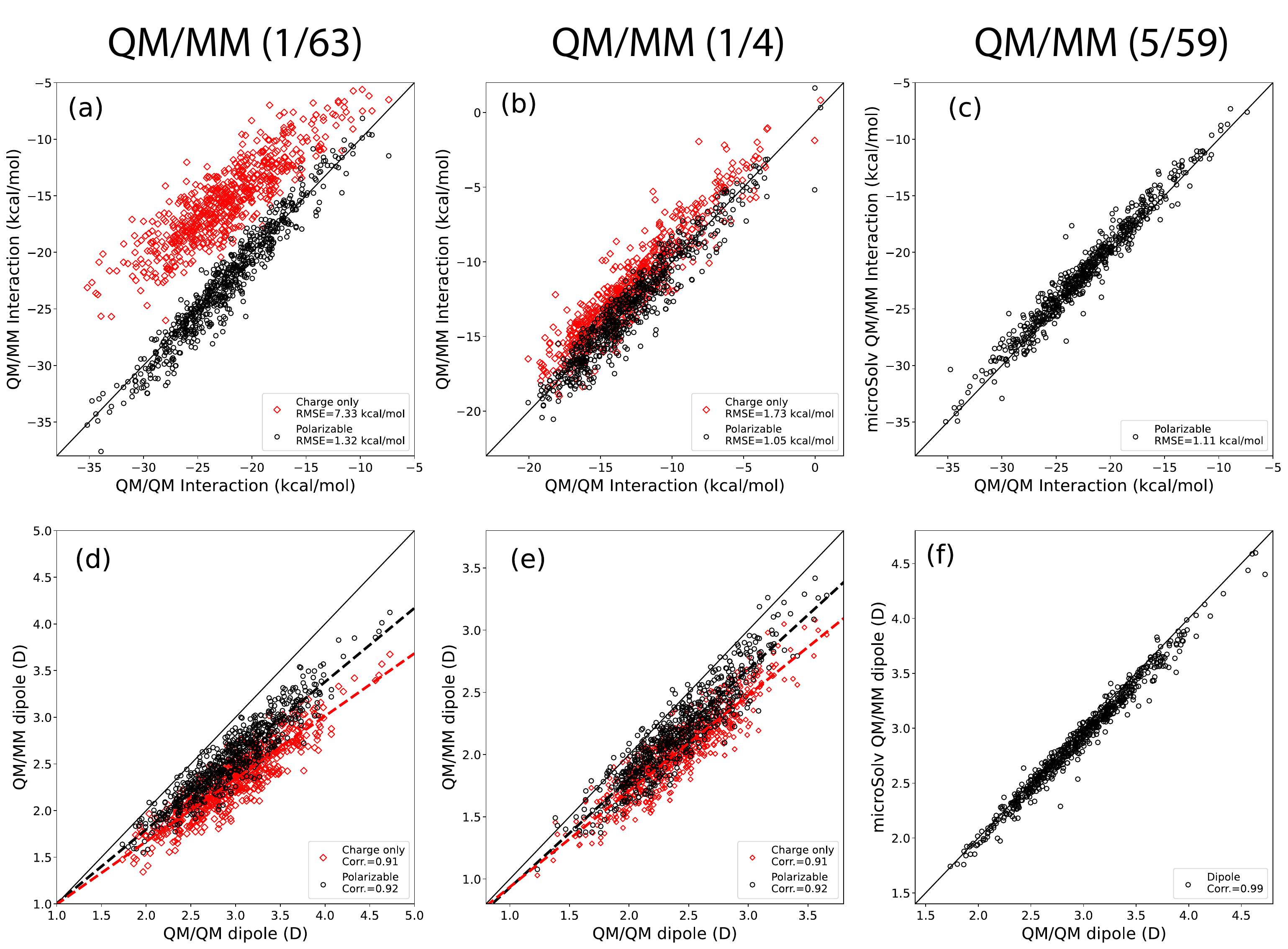}
  \caption{Panels (a), (b) and (c): correlation plots of the interaction energy (in kcal/mol) of a single water molecule with its environment in a model of liquid water. (a) Bulk: QM/MM with 1 QM water molecule and 63 MM water molecules. (b) First shell: 1 QM water molecule and 4 MM water molecules (only the first solvation shell). (c) Minimal solvation: 5 QM water molecules and 59 MM water molecules. Panels (d), (e), and (f): correlation plots of the dipole moment length of the embedded molecule (in Debye) for the same systems as for panels (a--c).}
  \label{fig:bulk_fs}
\end{figure}

Panels (a) and (b) of Figure \ref{fig:bulk_fs} show that the interaction energy between water molecules and their environment in liquid water ranges from -35 to -7 kcal/mol for the bulk calculations and from -20 to -1 kcal/mol for the first shell calculations. Such energy scales are in line with comparable calculations in the literature \cite{MHG_polQMMM}. As indicated in the figure, the root mean square errors (RMSEs) of the QM/MM interaction energy compared to the QM/QM ones are 1.32 kcal/mol for the bulk and 1.05 kcal/mol for the first shell calculations. \jm{The minimal solvation setup improves upon the bulk with an RMSE of 1.11 kcal/mol}. The RMSE of the first shell simulations compares well with 1.4 kcal/mol from a similar simulation from Ref.\ \citenum{MHG_polQMMM} where MM was treated with the AMOEBA force field. To our knowledge, a similar comparison for the bulk or \jm{minimal solvation} results is not available. 

\begin{figure}
  \centering
  \includegraphics[width=0.5\linewidth]{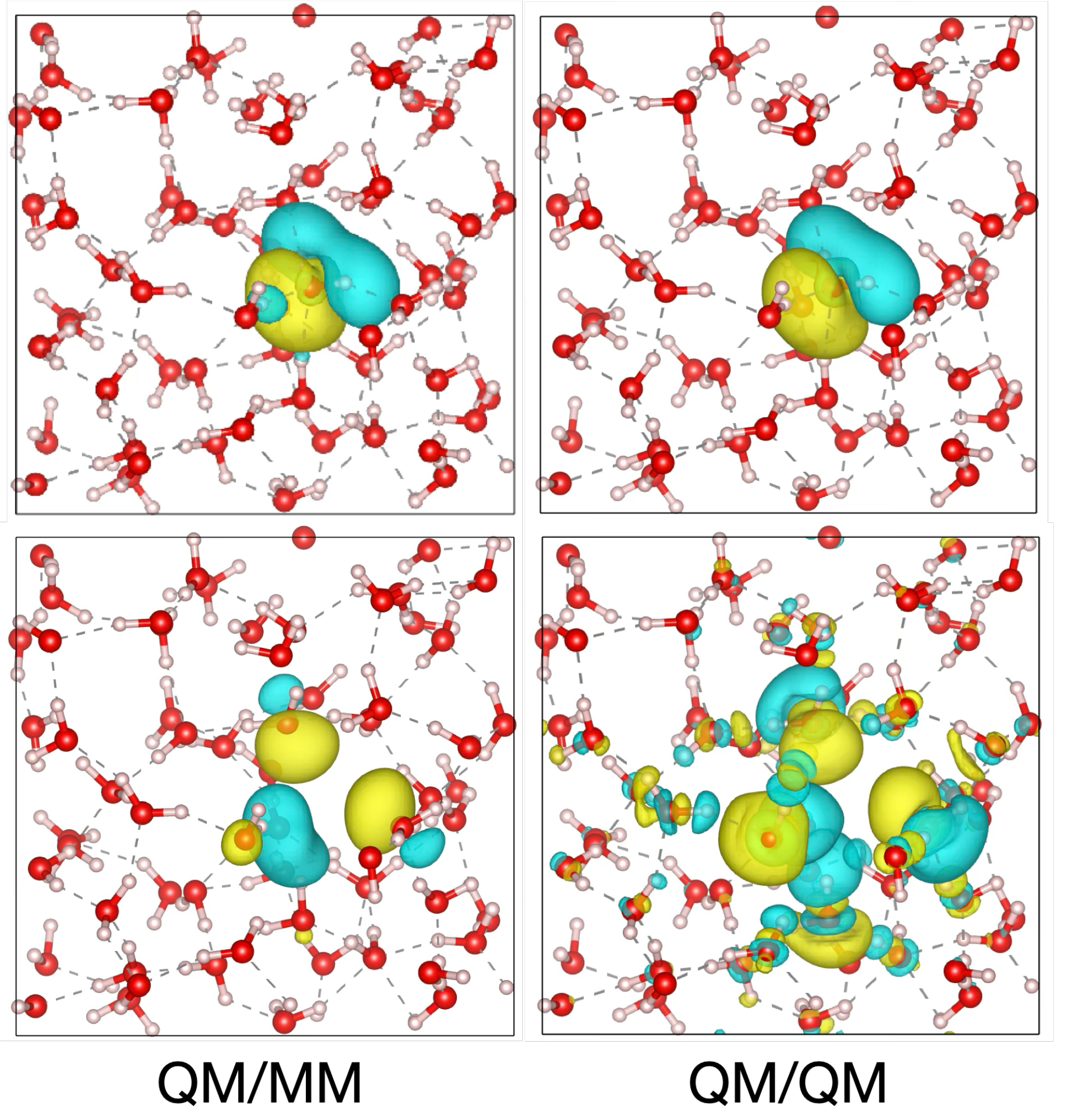}
  \caption{\label{bulkedd} Polarization density (defined as the difference of the embedded and isolated molecular electron density) of an embedded water molecule employing the methods indicated in the figure. Top panels: single water molecule polarization. Bottom panels: environment polarization. The isosurface value is set to $\pm 0.0025$ e$\cdot a_0^{-3}$.}
\end{figure}

Such RMSEs are also consistent with the accuracy of the hexamer structure discussed before. However, $\sim 1$ kcal/mol accuracy for bulk systems is even more impressive. In the hexamer, the interaction energies (mbpol) average about 8.00 kcal/mol per water molecule. In the bulk water, the QM--QM interaction energy averages to a much larger value ($\sim 22.42$ kcal/mol) due to a stronger polarization. In fact, the $\sim 1$ kcal/mol accuracy leads us to conclude that the QM-MM mutual polarization predicted by the QM/MM simulation \jm{in bulk and minimal solvation setups} is well represented. As we can see in Figure \ref{bulkedd}, both the polarization of the embedded water molecule (see panels a, b, and c) as well as of the environment (see panels e, f, and g) are fairly well represented by the QM/MM calculations. An interesting aspect of the MM polarization in panels (e) and (d) of Figure \ref{bulkedd} is that, in comparison to the QM/QM polarization they only feature a small component from the second solvation shell. An explanation for this effect is the fact that polarizable force fields typically employ slightly damped atomic dipole polarizabilities compared to the reference QM value \cite{tang1984improved,tkatchenko2013interatomic}. The damping is crucial to fend off overpolarization, which in our framework would stems from the neglect of the nonadditive components of the embedding potentials, see Eq.\ (\ref{eq:emb_MM}). In follow-up work, it will be interesting to analyze the effect of these additional terms on the MM polarization and the chosen values of the atomic dipole polarizabilities.

Panels (d) (e) and (f) of Figure \ref{fig:bulk_fs} present the dipole lengths of the central, embedded water molecule. These range from 1.6 D to 4.5 D for both 1/63 and 5/59 bulk simulations, and from 1.3 D to 3.5 D for the 1/4 first-shell calculations. Generally, we notice that the dipole lengths are underestimated in the QM/MM treatment compared to the reference QM/QM. The QM/MM dipoles, however, correlate very well with the reference QM/QM (sDFT) dipoles showing Pearson's correlation coefficients above 0.9. We also computed an RMSE for the dipoles of the 1/4 calculations of 0.29 D which compare well with the 0.62 D of Ref.\ \citenum{MHG_polQMMM}. The dipole results and the results from the interaction energies sustain our claim that our method is the most accurate QM/MM result for liquid water available to date.

Figure \ref{fig:bulk_fs} also shows (red diamonds) results labeled as ``charge only,'' which are from QM/MM simulations where the polarization of the MM subsystem is neglected (i.e., the polarizable dipoles are set to zero). While for the first shell systems the results are still acceptable (dipole correlation of 0.91, dipole RMSE of 0.42 D, and for the interaction energy of 1.73 kcal/mol), the results for the bulk system are qualitatively incorrect. In the bulk, the RMSE for the interaction energy jumps to 7.33 kcal/mol. An unacceptable deviation is, however, in line with the expected accuracy of a typical electrostatic embedding QM/MM simulation of condensed phases \cite{boereboom2016toward,chen2022improving}.

 \jm{When employing 63 coupled subsystems to represent the QM/QM benchmark, the interaction energies and dipole moment correlations are consistent with what is presented above, see supplementary Figure \ref{sup-fig:int_ener_63_subs}. We also note that MB-Pol provides us with a very accurate benchmark for the interaction energies. We thus include a comparison of the computed QM/MM interaction energies for the 1/63 system against an MB-Pol reference in supplementary Figure \ref{sup-fig:Bulk_MBPOL}.}

\subsection{Convergence with respect to the QM size}
\begin{figure}
  \centering
  \includegraphics[width=\linewidth]{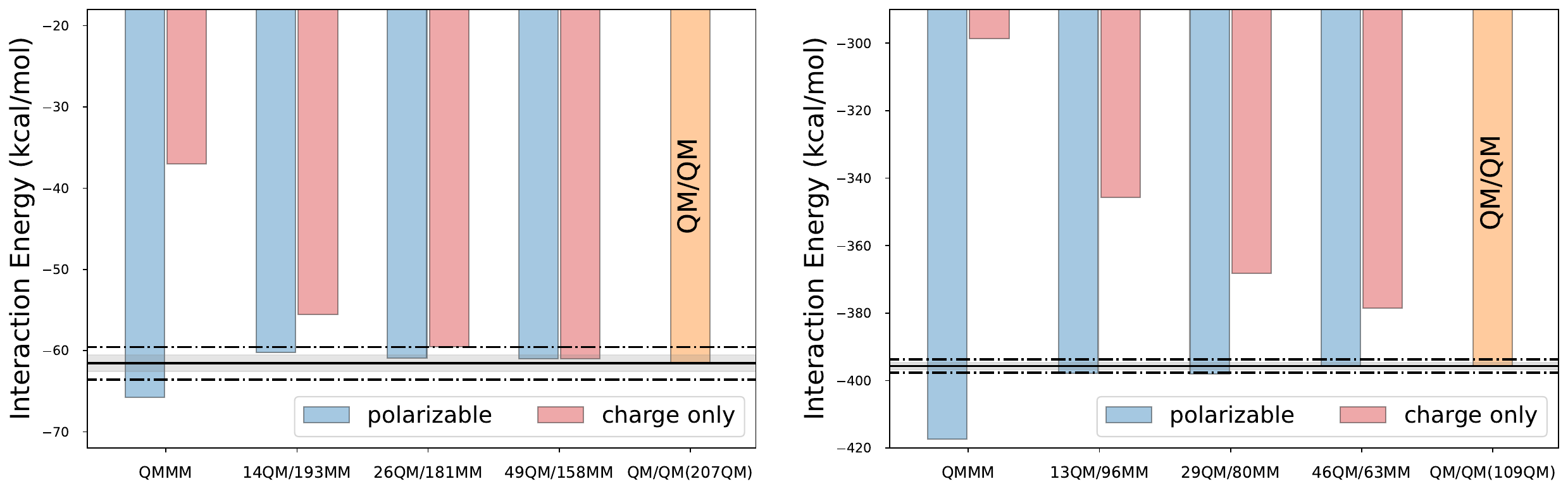}
  \caption{Convergence of the glucose (left panel) and the [Pd(H$_2$O)$_4$]$^{2+}$ aqua ion  (counterbalanced by 2 Cl$^-$ ions, right panel) solute-water interaction energy with respect to the number of water molecules included in the QM subsystem. An orange bar representing the sDFT reference is labeled as QM/QM in each plot. The dot-dashed line marks $\pm 2$ kcal/mol, and the shade marks $\pm 1$ kcal/mol window from the sDFT reference.}
  \label{fig:qm_conv}
\end{figure}

An important test for QM/MM simulations of solvated species is the convergence with respect to the number of water molecules included in the QM subsystem. As mentioned in the introduction, having an accurate model for the QM-MM interface and employing an MM force field that is consistent with the QM method (we use the MB-PBE force field) should allow us to showcase strong QM/MM convergence. The typical target is a $\pm$ 2 kcal/mol from the reference QM calculation \cite{perez2023effect}. In Figure \ref{fig:qm_conv}, we present the interaction energies of glucose (left panel) and the [Pd(H$_2$O)$_4$]$^{2+}$ aqua ion, counterbalanced by 2 Cl$^-$ ions (right panel), with the water bulk environment as a function of the number of water molecules included in the QM subsystem. For glucose (a neutral molecule), the figure clearly shows that the $\pm 2$ kcal/mol target is reached (in fact, a $\pm 1$ kcal/mol is achieved) already when only 14 water molecules are included in the QM subsystem. This amounts to including less than the first solvation shell. Interestingly, the convergence for this system is much worse when a charge-only MM model is used (i.e., as done before, we simply neglect the polarization of the MM subsystem) for which 26 QM water molecules are needed to reach convergence. For PdCl$_2$, despite the presence of a doubly charged cation, we find an essentially identical behavior with convergence being reached already with 13 QM water molecules. Conversely, the charge only QM/MM method does not reach the $\pm 2$ kcal/mol goal even when 46 QM water molecules are included.  Therefore, we conclude that similar to the glucose system, including polarization in the MM subsystem dramatically improves the convergence of the interaction energy with the size of the QM subsystem.

Our results on the need to include polarization in the MM subsystem find justification in the fact that, for a condensed phase system, including the inductive response of the environment, is crucial to obtain a physical picture, particularly for charged solutes \cite{MHG_polQMMM}.

In supplementary Figure \ref{sup-fig:glucose-pol}, we also present the polarization density of glucose and its water environment, showing once again that the polarization of the QM/MM model reproduces fairly accurately the polarization of the reference QM/QM simulation.

\subsection{Wet Surfaces}

In the supplementary information document \cite{epaps}, we also present a QM/MM calculation of a monolayer MoS$_2$ solvated by water, see Figures \ref{sup-fig:mos2-scan} and \ref{sup-MoS2pol}. The conclusions of that analysis are consistent with what has been presented so far. That is, QM/MM simulations reproduce within a reasonable error the QM/QM result for the interaction energy. For this system, we record a QM/MM interaction energy of \jm{-0.11 kcal/mol $\cdot$ \AA$^{-2}$ at the equilibrium distance compared with a QM/QM reference of -0.23 kcal/mol $\cdot$ \AA$^{-2}$, and QM/MM polarization densities very close to the QM/QM benchmark.} The equilibrium water-MoS$_2$ distance of 2.1 \AA\ is predicted by both QM/QM and QM/MM methods.

\section{Conclusions}

We introduced a novel QM/MM framework that incorporates density-functional theory (DFT) for the QM and the MM subsystems that leverages an orbital-free treatment for the MM region and the QM-MM interaction. By assigning an electron density to the MM subsystem and accurately capturing nonadditive interactions (such as exchange, correlation, Coulomb, and Pauli repulsion effects) our density-functionalized QM/MM approach achieves chemical accuracy in modeling complex solute-solvent systems. We validated the approach against a variety of water-based systems, including water clusters, bulk water, solvated ions, and a wet monolayer of MoS$_2$, demonstrating consistent accuracy and achieving unprecedented fast convergence to chemical accuracy with respect to increasing size of the QM subsystem. Reaching chemical accuracy with the proposed method requires only the inclusion of the first solvation shell in the QM region--a significant advancement over traditional QM/MM schemes.

Particularly striking is the performance of our density functionalized QM/MM method for H$_2$O in water. There we find that the excellent performance of the method for clusters (dimers, pentamers and hexamers) seamlessly translates to accurate models of bulk liquid water. A prime example is given by the dipole of the solute molecule which we predict an RMSE with QM/MM of 0.29 D, or 12\%, for both bulk and pentamers compared to 0.624 D for pentamers from Ref.\ \citenum{MHG_polQMMM}.

Our results highlight the critical role of (1) employing ab-initio density functionals for the QM-MM interactions instead of ad-hoc parametrizations; (2) properly accounting for mutual polarization at the QM-MM interface; and (3) employing MM force fields that are consistent with the QM method employed (here we use MB-PBE for MM and DFT with the PBE exchange-correlation functional for QM). We found that following these three principles significantly improves convergence with respect to the size of the QM region compared to standard QM/MM methods. Our pilot simulations have showed that for both neutral and charged solutes, the interaction energies reached the target accuracy of within $\pm$ 2 kcal/mol using a minimal QM region, with further refinement yielding sub-1 kcal/mol errors when merely the first shell of solvent molecules is included in the QM subsystem. This level of accuracy, achieved even in bulk water systems, underscores the robustness of our method for simulating condensed-phase environments.

Overall, this work presents a significant step forward in extending the applicability of QM/MM methods to treat larger, more complex systems than typically approached by standard QM methods while maintaining chemical accuracy. Future work will explore the impact of including beyond-Coulomb, ab-initio terms in the MM embedding potential, so that the MM dipole response can more closely resemble the true electronic response of the electrons in the MM subsystem. We plan to apply the density-functionalized QM/MM framework to chemical environments other than water, for example those provided by biomolecules and materials interfaces as well as non-aqueous solvents.

\section*{Acknowledgements}
This research was partially funded by the U.S.\ National Science Foundation grants No.\ CHE-2154760 (MP, JMB and XC), OAC-2321103 (MP, JMB and XC), OAC-2311260 (MR and FP) and CHE-2306929 and OAC-2321102 (OA). All computations were carried out on the Price supercomputer of Rutgers University-Newark acquired through an NSF MRI grant No.\ OAC-2117429 (MP) and managed by the Office of Advanced Research Computing at Rutgers.

\section*{Authors contributions}
Pavanello, Andreussi and Paesani conceived the work
and co-wrote the manuscript. Pavanello managed the
students and the project. Shao and Riera developed
the software. Martinez B. and Chen carried out the
simulations and co-wrote the manuscript.

\providecommand{\latin}[1]{#1}
\makeatletter
\providecommand{\doi}
  {\begingroup\let\do\@makeother\dospecials
  \catcode`\{=1 \catcode`\}=2 \doi@aux}
\providecommand{\doi@aux}[1]{\endgroup\texttt{#1}}
\makeatother
\providecommand*\mcitethebibliography{\thebibliography}
\csname @ifundefined\endcsname{endmcitethebibliography}
  {\let\endmcitethebibliography\endthebibliography}{}

\end{document}